# Real-space imaging of polar and elastic nano-textures in thin films via inversion of diffraction data

Ziming Shao[1], Noah Schnitzer[1], Jacob Ruf[2], Oleg Y. Gorobtsov[1], Cheng Dai[3], Berit H. Goodge[4,5], Tiannan Yang[3], Hari Nair[1], Vlad A. Stoica[3,6], John W. Freeland[6], Jacob Ruff[7], Long-Qing Chen[3], Darrell G. Schlom[1,5,8], Kyle M. Shen[2,5], Lena F. Kourkoutis[4,5], Andrej Singer[1*]

[1]Department of Materials Science and Engineering, Cornell University, Ithaca, NY 14853, USA

[2]Department of Physics, Cornell University, Ithaca, NY 14853, USA

[3]Department of Materials Science and Engineering, Pennsylvania State University, University Park, PA 16802, USA

[4]School of Applied and Engineering Physics, Cornell University, Ithaca, NY 14853, USA

[5]Kavli Institute at Cornell for Nanoscale Science, Cornell University, Ithaca, NY 14853, USA

[6]Advanced Photon Source, Argonne National Laboratory, Argonne, IL 60439, USA

[7]Cornell High Energy Synchrotron Source, Cornell University, Ithaca, NY 14853, USA

[8]Leibniz-Institut für Kristallzüchtung, Max-Born-Straße 2, 12489 Berlin, Germany

*asinger@cornell.edu

**Introductory paragraph:**

**Exploiting the emerging nanoscale periodicities in epitaxial, single-crystal thin films is an exciting direction in quantum materials science: confinement and periodic distortions induce novel properties. The structural motifs of interest are ferroelastic, ferroelectric, multiferroic, and, more recently, topologically protected magnetization and polarization textures. A critical step towards heterostructure engineering is understanding their nanoscale structure, best achieved through real-space imaging. X-ray Bragg coherent diffractive imaging visualizes sub-picometer crystalline displacements with tens of nanometers spatial resolution. Yet, it is limited to objects spatially confined in all three dimensions and requires highly coherent, laser-like x-rays. Here we lift the confinement restriction by developing real-space imaging of periodic lattice distortions: we combine an iterative phase retrieval algorithm with unsupervised machine learning to invert the diffuse scattering in conventional x-ray reciprocal-space mapping into real-space images of polar and elastic textures in thin epitaxial films. We first demonstrate our imaging in PbTiO₃/SrTiO₃ superlattices to be consistent with published phase-field model calculations. We then visualize strain-induced ferroelastic domains emerging during the metal-insulator transition in Ca₂RuO₄ thin films. Instead of homogeneously transforming into a low-temperature structure (like in bulk), the strained Mott insulator splits into nanodomains with alternating lattice constants, as confirmed by cryogenic scanning transmission electron microscopy. Our study reveals the type, size, orientation, and crystal displacement field of the nano-textures. The non-destructive imaging of textures promises to improve models for their dynamics and enable advances in quantum materials and microelectronics.**

**Main text:**

A century after its discovery, X-ray diffraction is the pillar of the structural characterization of crystalline materials. The diffracted X-ray intensities, $I(\vec{Q})$, relate to the density of scatterers, $\rho(\vec{r})$, through a Fourier transform $I(\vec{Q}) = A\left|\int \rho(\vec{r}) \exp(-i\vec{Q} \cdot \vec{r}) d\vec{r}\right|^2$, where $\vec{Q}$ is the momentum transfer in reciprocal space, $\vec{r}$ is the real-space coordinate, and $A$ is a constant. In periodic structures, x-ray interference generates intensity maxima (Bragg peaks), whose position, relative intensities, and shape allow studying the structure of the crystal unit cell, crystal lattice, mesoscale periodicities, and interfaces such as twin and antiphase domain boundaries[1]. The Fourier transform is invertible, yet the notorious phase problem – one can only measure x-ray photon intensities, not their phases – prevents a direct inversion of the x-ray data into a real-space structure. Protein crystallography offers powerful approaches for solving the structure directly, for example, using multiple measurements around an x-ray absorption edge in multi-Wavelength Anomalous Diffraction (MAD)[2,3]. Apart from a few exceptions, the analysis of x-ray data entails refining a model by comparing its calculated diffraction intensity with the measured one[4]. While protein crystallography relies on a large database of know structures, refining extended solid-state structures is more challenging due to the lack of a database and often requires sophisticated phase-field modeling.

As proposed by Sayre[5,6], diffraction data can be inverted if intensities between Bragg reflections are recorded. Sayre's idea culminated in the development of coherent x-ray diffractive imaging (CXDI)[7–13]: a lensless imaging technique with the resolution in principle limited only by the wavelength and the strength of the scattering signal. Akin to laser diffraction on a pinhole, a coherent x-ray beam scattered off a spatially confined specimen produces sharp interference fringes. These fringes can be inverted into a real-space image if the entire specimen is illuminated with a spatially coherent, laser-like x-ray beam and the specimen size is small enough for the interference fringes (spaced inversely proportional to the specimen size) to be measured with sufficient resolution[14,15]. CXDI has been used for visualizing biological cells[16,17], strain in nanocrystals[9,18], and recently operando changes in energy materials[19,20] and single macromolecules at free-electron lasers[21,22].

Interestingly, interference fringes can also emerge around a Bragg reflection of single-crystalline thin films as 'satellite peaks' when a periodic nano-texture is present[23–26]. These satellite peaks occur even when measured on large samples with an X-ray beam of poor spatial coherence. The satellite peaks emerge due to the highly periodic nano-texture, reminiscent of protein crystals, where the unit cell is no longer a protein but a periodic unit cell of lattice distortions. Consider an arrangement of structural motifs, which displays short-range periodicity over small areas (see Fig. 1) but lacks the long-range order due to domain walls. Despite the lack of long-range periodicity and confinement in the horizontal direction ($x$ in Fig. 1), the diffracted intensity, $I(\vec{Q})$, of such a structure displays well-defined satellite peaks, in addition to the sharp central Bragg peak and thickness fringes due to confinement in the vertical direction ($z$ in Fig. 1). The satellite peaks persist even with slight distortions in the short-range order (note the differing distance between the triangles from spot to spot in Fig. 1). Here, we combine conventional iterative phase retrieval[27] with unsupervised machine learning[28,29] to extract high-resolution images of the structural motifs (Eigen Patterns) from the diffraction intensities, $I(\vec{Q})$, measured from epitaxial thin crystalline films (such as one shown in Fig. 1). The method is akin to crystallography: we reveal the structure of the supercell (mesoscale unit cell) by using its periodic arrangement within the superlattice from a single reciprocal space map. Nanobeam diffraction has been used to extract strain in thin films in 2D and 3D.[30,31] Our method produces higher-resolution images and requires no nano-focusing or spatial scanning, significantly reducing data acquisition time and thereby enabling the investigation of dynamics at fast timescales.

Thin epitaxial films are extensively used for investigating the fundamental interactions in solids and for microelectronics. Strain engineering through heteroepitaxial growth is a way to control the electronic, magnetic, or structural properties of quantum materials[32], and can introduce scientifically and technologically critical structural motifs at the mesoscale.[24,30,33] The recent advances in atomic deposition techniques allow synthesizing high-quality crystals of complex oxides, which generate stunning diffraction patterns resembling those measured with optical lasers or focused coherent x-ray beams from single-crystalline particles in Bragg coherent diffractive imaging[9]. One such system is the multilayer $(PbTiO_3)_n/(SrTiO_3)_n$ [34], with an optically induced coherent 3D supercrystal[23]. The superlattice is revealed by the X-ray diffraction pattern around the 004 $DyScO_3$ substrate (DSO) pseudo-cubic Bragg peak of $(PbTiO_3)_{16}/(SrTiO_3)_{16}$ multilayer system, in which evenly spaced off-specular satellite peaks were observed within the reciprocal cross-section along $Q_z\|[110]_{DSO}$ and $Q_x\|[-110]_{DSO}$ ( see Fig. 2a). Due to the Fourier slice theorem, the shown cross-section $I(Q_x,Q_z)$ is the Fourier transform of the projection of the structure in the $Q_y$ direction: it represents a cross-sectional view of the film along the substrate surface in the $[001]_{DSO}$ direction.

To invert the diffraction data, $I(Q_x, Q_z)$, we use a conventional 2D phase retrieval algorithm (see methods)[27]. Figure 2b shows the calculated squared magnitude of the Fourier transform of the structure, retrieved through an iterative algorithm from data (see Fig. 2a) initiated with a fully randomized guess. The reconstruction successfully reproduces the main features of the measured diffraction data, such as the satellite peaks and the Laue fringes in between (compare Figs. 2a,b). Akin to Bragg coherent diffractive imaging[9], the retrieved image $\rho(x,z) = s(x,z)\exp[i\,\vec{Q}\cdot\vec{u}(x,z)]$, consists of the electron density of Bragg planes, $s(x,z)$, and the displacement field projected along the measured reciprocal space vector, $\vec{Q}\cdot\vec{u}(x,z) = Q_z u_z(x,z)$. We treat the structure as a pure phase object with $s(x,z) = 1$. Figure 2c shows the retrieved displacement field, $u_z(x,z)$, (along $Q_{004}$) used to calculate X-ray data in Fig. 2b. A 2D periodicity is apparent. Moreover, calculating the derivative of the vertical displacement field $u_z(x,z)$ along the vertical direction, $z$, yields normal strain (NS), and along the horizontal direction, $x$, it yields the local Crystal-plane inclination (CI) of the lattice planes (see methods). The crystal-plane inclination and normal strain maps (see Fig. 2d,e) reveal the object to be a stack of alternating horizontal layers with a half-period offset in the horizontal direction.

A critical consistency test of phase retrieval is the averaging over multiple runs of the algorithm initiated by random starting conditions[10,14,27,35,36]: a reconstruction is successful when the squared modulus of the Fourier transform of the average real-space image closely resembles the measured intensity, which requires all runs of the algorithm to converge to a unique real-space solution. This requirement is not fulfilled here. All real-space images display patches of short-range-ordered 2D checker-board texture; nevertheless, the area and position of periodic regions are inconsistent across the differently initialized phase retrieval runs, preventing us from efficient averaging (only 100 out of 2000 individual reconstructions display a high mutual correlation, see Fig. S1). Indeed, the variation in the long-range order is unsurprising: the data collected with partially coherent, unfocused x-rays lacks the interference between the multiple illuminated areas on the film (either due to the low spatial x-ray coherence or the interference fringes being much smaller than the detector pixel size).[37]

Despite the failure of the conventional averaging, we extract the characteristic structural motifs of the short-range order by applying unsupervised machine learning to the set of reconstructed structures. First, we define a periodic mesoscale 'unit cell', the supercell, and determine its period in $x$ and $z$ direction

$(p_x, p_z)$ from the satellite peak positions in the diffraction data. We determine the supercell-locations by cross correlating the CI with a test function, $\sin(2\pi x/p_x) \cdot \sin(2\pi z/p_z)$. We then partition $2*10^3$ images of CI (see Fig. 2d) into an ensemble of $2.2*10^4$ supercells of size $p_x \cdot p_z$. Finally, we classify and label the supercells through k-means clustering and call the resulting structural motifs Eigen Patterns (EP) (see Figs. S2, S3). Figure 2f-g shows the CI and the NS maps of the most dominant eigen pattern. As expected from the synthesis, the normal strain shows the bilayer with a larger (smaller) lattice constant of $PbTiO_3$ ($SrTiO_3$). The Crystal-plane inclination reveals patches of up tilt (red) and down tilt (blue), when following the Bragg planes from left to right. The imaging is consistent with the phase-field modeling presented in the original publication [23] and reveals additional features, including the impact of the plane inclination on the lattice constant (tensile strain between tilt-up and tilt-down regions in the $SrTiO_3$ layer) (see Fig. S4).

Having demonstrated the combination of k-means with phase retrieval on a previously studied system, we use it to investigate the impact of strain on the Mott transition in $Ca_2RuO_4$ – an important member of the strongly correlated materials family with a wide variety of electronic and magnetic phenomena. Notably, $Ca_2RuO_4$ features a Mott metal-insulator transition, which strongly couples to a structural transition from low-temperature S-Pbca to high-temperature L-Pbca phases[38–41]. $Ca_2RuO_4$ thin films have been successfully grown upon various substrates, and the metal-to-insulator transition temperature has been tuned by biaxial in-plane epitaxial strain[42–44]. When grown on $LaAlO_3$ (LAO), instead of the abrupt resistivity change at $T_{MIT}$=357K observed in bulk $Ca_2RuO_4$, the metal-insulator transition broadens in temperature, and the transition temperature reduces to 230 K, revealed by the change in sign of dR/dT (resistance (R), temperature (T), see Fig. 3a). The suppressed metal-insulator transition temperature has been ascribed to the compressive epitaxial strain, which impedes the expansion of the $Ca_2RuO_4$ lattice within the *ab* basal plane during the accompanying structural transition[42] (see Fig. 3b).

To achieve a better understanding of the structural behavior during the Mott transition in strained 14 nm thick $Ca_2RuO_4$ films epitaxially grown on $LaAlO_3$ substrate, we collected X-ray diffraction reciprocal space maps around the 008 Bragg peak at 300 K (metal) and 7 K (insulator) (see Fig. 3c). Above the metal-insulator transition temperature, a sharp Bragg peak with Laue fringes indicates the excellent quality of the $Ca_2RuO_4$ thin film. Below the metal-insulator transition temperature, additional satellite peaks emerge perpendicular to $Q_z \| [001]_{LAO}$. The satellite peaks are confined to two orthogonal planes $Q_z$-$Q_x$ and $Q_z$-$Q_y$, where $Q_x \| [110]_{LAO}$ (Fig. 3d(i-iii)). The line profile along $Q_x$ (or $Q_y$, not shown) at Q =006, 008, and 0014 display evenly spaced satellite peaks with a position in $Q_x$ independent of the Bragg peak (see Fig. 3e(i-iii)), indicating nanoscale periodicity in the $Ca_2RuO_4$ film as the origin of satellite peaks instead of lattice tilting. A tilting of large domains would lead to larger satellite distances in $Q_x$ at higher momentum transfer $Q_z$.[45] We estimate the corresponding periodic length as $2\pi/\Delta q \approx 17.5$ nm ($\Delta q$=distance between satellite peaks along $Q_x$), which is temperature independent (see Fig. S5). Moreover, we observe a peak broadening by comparing the 1st, 2nd, and 3rd order satellite peaks in the $Q_x$ direction, suggesting slight period length variations between different patches of short-order domains.

The diffraction pattern of the 008 peak at 7 K in the $Q_x$-$Q_z$ plane shows high-quality interference fringes with multiple satellites present in both directions (see Fig. 4a), suggesting an intricate superlattice structure. Because all satellite peaks are confined to two distinct planes ($Q_x$-$Q_z$ or $Q_y$-$Q_z$), we conclude that they emerge from two spatially distinct types of heterogeneities, oriented perpendicularly to each other (indistinguishable by the symmetry of the substrate, see Fig. S6). Diffraction confined to the $Q_x$-$Q_z$ plane suggests nanoscale heterogeneity in the x-z plane, extended over a larger length in the y-direction. The diagonal streaks in the diffraction data in the $Q_x$-$Q_z$ plane further indicate structural motifs with diagonally oriented boundaries in the real-space structure in the x-z plane.

The typical phased diffraction data from a randomized starting guess retains most of the features in the measured peak, including the diagonally placed satellite peaks and their relative intensities (compare Figs. 4a and 4b). The displacement field of the corresponding real-space structure exhibits a continuous periodic zig-zag-shaped modulation (see Fig. S7). The zig-zag pattern comprises diagonally oriented domain-wall-like structures consistent with diagonal streaks in the diffraction pattern. Inspecting the individual phase retrieval reconstructions with the random initialization, one notices the presence of a similar in-plane periodicity with various types of supercells, all reproducing the measured diffraction pattern (see Fig. S7). Yet only 200 out of 2000 reconstructions display a high correlation, again preventing us from conventional averaging.

By clustering the individual supercells, we identify four distinct EPs of supercells composing the short-range order (see Fig. 4c,d and methods). The features are well resolved, though only 18 nm large with a real-space image pixel size of ~2 nm * 1 nm. EP1 and EP2 correspond to the individual supercell of the zig-zag pattern we saw in the average (see Fig. S7), showing a triangular shape structure with a smaller *c*-axis lattice constant in the NS map and a cross shape with opposite tilting of the (008) lattice planes along the two diagonals in the CI map. EP3 and EP4 both show a smaller *c*-axis lattice constant diagonal stripe in the NS map (see Fig. 4c(iii-iv)) and a diagonal stripe in the corresponding CI map (see Fig. 4d(iii-iv)). The directions of the diagonals are opposite between EP3 and EP4 but consistent in the NS and CI for the same EP. The population ratio collected from the clustering of the four EPs is ~2:2:1:1. Because the EP1 and EP2 supercells are the majority in the reconstructions, the appearance of the zig-zag pattern in the average result (see Fig. S7) is statistically plausible as EP1 and EP2 get a better mutual correlation. In the aspect of symmetry, EP1 and EP2 present a reflection symmetry within themselves, producing symmetry in the diffraction pattern around the vertical axis. Although EP3 and EP4 lack reflection symmetry, they are mirror reflections of each other, and their similar populations generate mirror symmetry in the diffraction pattern. A possible explanation of the presence of both EP1 and EP2 is Friedel's law, which prevents us from distinguishing the up from the down direction. Nevertheless, we never saw both EP1 and EP2 in a single reconstruction.

To confirm the validity of our x-ray data inversion, we performed cryogenic high-angle annular dark-field (HAADF) scanning transmission electron microscopy (STEM) at ~100K on the cross-section of a similar ~34nm thick $Ca_2RuO_4$/$LaAlO_3$ film. Figure 4e shows a map of the local interplanar spacing along the $[001]_{LAO}$ direction extracted from the atomic resolution HAADF-STEM images, which confirms the presence of mirrored diagonal features similar to those visible in the reconstructed EP3 and EP4. Some mirrored diagonals overlap, forming triangular patterns similar to EP1 and EP2 in our reconstruction from the x-ray data. In addition to the similar morphology, the average angle of the diagonal in the STEM measurement is ~48 degrees, which matches the angle calculated from our x-ray imaging (see Fig. 4d(iv)). Although the $Ca_2RuO_4$ film shows local mesoscale periodicity, variations in the short-range order are present, and a long-range regular strain ordering is not observed within the STEM measurement.

The striped nano-texture of the epitaxially strained $Ca_2RuO_4$ thin film revealed by the real-space imaging likely results from elastic energy minimization, similar to the formation of nano-textures in other complex oxides[23]. The nanotexture disappears at high temperatures above the Mott transition and reappears at low temperatures in the Mott state, creating an exciting opportunity for controlling the nano-texture by

Mott physics. The nano-texture has not been reported in bulk $Ca_2RuO_4$, suggesting epitaxial strain stabilizes it. The out-of-plane lattice constant of alternating stripes indicates that the two structures are derivatives of the high-temperature metallic and low-temperature insulating phases. Nevertheless, spatially resolved studies are needed to better understand the correlation of the observed structural texture with charge, spin, and orbital ordering, such as near-field optical nanoscopy[33], resonant scanning X-ray microscopy, or energy-resolved electron microscopy. An exciting future direction is studying and understanding the femtosecond to picosecond dynamic evolution of the nano-texture in Mott insulators such as $Ca_2RuO_4$ because of the new excitations and new timescales that can form due to the new periodicity.

In summary, we demonstrated the real-space imaging of the periodic structural nano-texture in $(PbTiO_3)_{16}/(SrTiO_3)_{16}$ and $Ca_2RuO_4$ epitaxial thin films through the combination of Bragg Coherent Diffractive Imaging and unsupervised machine learning. Our model-independent x-ray imaging approach agrees with the phase-field modeling in $(PbTiO_3)_{16}/(SrTiO_3)_{16}$ and cryo-STEM measurement results in $Ca_2RuO_4$. The real-space images of nano-texture motifs represent averages over millimeter-large, multi-domain areas. The success of the approach, therefore, requires a nanotexture with uniform periodicity over extended areas. Nevertheless, the recent developments in atomic deposition technologies make the technique highly applicable in visualizing the ubiquitous nano-textures in low-dimensional quantum materials and microelectronics, often displaying so-called "butterfly" or more complex diffraction patterns.[23–26] The averaging over extensive areas highlights our technique's complementarity with local imaging probes such as Bragg coherent diffractive imaging, ptychography, and atomic-resolution electron microscopy, all limited to relatively small film areas and potentially prone to local defects. The method shows some similarity with Coherent Bragg Rod Analysis (COBRA).[46] While COBRA requires interference between the film and the substrate scattering to determine the atomic positions within an average unit cell, our method retrieves the nano-texture from diffuse scattering from the films. The method can therefore be applied to study strain induced nano-textures in free-standing membranes synthesized, for example, by remote epitaxy.[47] The combination of conventional phase retrieval and machine learning is a potential pathway to invert data with non-unique solutions, and we anticipate that the combination of our approach with more sophisticated phase retrieval algorithms, for example, using the multi-modal approach[15,48], will allow studying more complex systems with coexisting orders. Finally, the extension of our non-destructive imaging technique to in-situ and ultra-fast experiments at X-ray free-electron lasers appears straightforward.

**Acknowledgments:**

The work was primarily supported by U.S. Department of Energy, Office of Science, Office of Basic Energy Sciences, under Contracts No. DE-SC0019414 (development of the phase retrieval algorithm, data analysis and interpretation: Z.S., O.G., D.G.S., K.S., A.S.; thin film synthesis: H.N.). We thank Benjamin Gregory for careful reading of the manuscript. Research conducted at CHESS was supported by the National Science Foundation under awards DMR-1332208 and DMR-1829070. Use of the Advanced Photon Source was supported by DOE's Office of Science under contract DE-AC02-06CH11357. Transmission electron microscopy work was supported by the National Science Foundation (Platform for the Accelerated Realization, Analysis, and Discovery of Interface Materials (PARADIM)) under Cooperative Agreement No. DMR-2039380 and made use of the Cornell Center for Materials Research Shared Facilities, which are supported through the NSF MRSEC program (DMR-1719875). N.S. was supported by the NSF Graduate

Research Fellowship (DGE-2139899). B.H.G. was supported by PARADIM (NSF DMR-2039380). L.F.K. acknowledges support by the Packard Foundation. T.N.Y. and L.Q.C.'s efforts on phase-field simulations are supported as part of the Computational Materials Sciences Program funded by the U.S. Department of Energy, Office of Science, Basic Energy Sciences, under Award No. DE-SC0020145. C.D.'s effort is supported by the U.S. Department of Energy, Office of Science, Office of Basic Energy Sciences, under Award Number DE-SC-0012375. V.A.S., J.W.F., and L.Q.C. acknowledge the U.S. Department of Energy, Office of Science, Office of Basic Energy Sciences, under Award Number DE-SC-0012375 for support studying complex-oxide heterostructure with X-ray scattering.

## Methods

**Sample preparation:** The $Ca_2RuO_4$ thin film was grown in a Veeco Gen10 molecular-beam epitaxy (MBE) system on a $(001)_{pc}$-oriented $LaAlO_3$ substrate from CrysTec GmbH. The film was grown at a substrate temperature of 870 °C as measured using a pyrometer operating at 1550 nm. Elemental calcium (99.99% purity) and elemental ruthenium (99.99% purity) were evaporated from a low-temperature effusion cell and an electron beam evaporator, respectively. The films were grown with a calcium flux of $1.8\times10^{13}$ atoms·cm$^{-2}$s$^{-1}$ and a ruthenium flux of $1.7\times10^{13}$ atoms·cm$^{-2}$s$^{-1}$ in a background of $7\times10^{-7}$ Torr of ozone (10% $O_3$ + 90% $O_2$). At the end of the growth the shutter on both calcium and ruthenium sources were closed and the sample was cooled down to 250 °C in the same background pressure as used during the growth.

**Resistivity measurement:** The resistance is measured with the standard four-probe method, with four electrical contacts attached to the thin film near the corners of a 10 mm x 10 mm wafer. The square edges of the wafer are aligned along the [100] and [010] pseudo-cubic directions of $LaAlO_3$. Resistance is converted to resistivity using van der Pauw's formula for a 16.9 nm-thick conducting sheet.

**X-ray Diffraction (XRD):** The PTO/STO XRD data is taken from Stoica *et al.*[23] Detailed PTO/STO XRD experimental specifications can be found in the cited work. The $Ca_2RuO_4$ synchrotron XRD data was measured at A2 beamline of the Cornell High Energy Synchrotron Source (CHESS) with the incident photon energy of 19.75keV (wavelength $\lambda = 0.6278$ Å). The beam is unfocused during the diffraction measurement with the approximate illuminate size of ~$200\mu m * 400\ \mu m$ on the film. A Pilatus3 300K-pixel area detector ($487 \times 689$ pixels in an $83.8mm \times 106.5mm$ active area) was positioned 574.34 mm from the sample. The 3D reciprocal space around Bragg peaks of interest was acquired by $\theta$-scan (rocking-curve) of the sample, rotating in 320 steps of 0.0125 degree with one image recorded per step. Acquisition time is 1 second with an average of $2.7 \times 10^5$ photons collected for each image.

**2D Phase Retrieval Algorithm and Strain imaging:** The phase retrieval algorithm implemented in this work is a custom version of the Fienup's output-output (OO) algorithm[27]. (Code posted in SI) For each reconstruction initialized with a random start, we will run 20 mini-iterations, which is composed of 60 OO[0.6] iterations (output-output algorithm with feedback parameter $\beta = 0.6$), then 20 OO[0.8] iterations followed by 20 OO[0.98] iterations in the end. Also, we assume the object we reconstructed is a phase object (i.e., the object amplitude, $s(x,y)$, is constant). We assume the slight modulations in the amplitude intensity to be insignificant for the phase retrieval: in far-field diffraction, the phase information has more contribution to the resulting diffraction pattern compared to the amplitude information of the object.[13] A finite rectangular support is used as the real-space constraint. The vertical size of the support is determined by the thickness of the film, determined from thickness-fringes in the diffraction pattern. By fine-tuning the vertical support size and comparing the thickness fringes of the corresponding reconstructions with the measured pattern, the optimal vertical support size calculated is 75 nm for PTO/STO and 14 nm for $Ca_2RuO_4$. Compared to the vertical support size, the horizontal support size has less influence on the reconstruction result. And the reconstruction works best if the horizontal support size matches the integer multiple of the period length (supercell size). The horizontal support size selected in this work is the maximum size allowed without violating the oversampling principle, which is 172 nm for PTO/STO and 170 nm for $Ca_2RuO_4$ reconstruction (horizontal support introduces intensity modulations in the horizontal direction; at least two detector pixels per modulation are required for oversampling. See Fig. S10).

**Unsupervised machine learning for determining the supercell:** We first determine the size of the supercell according to the mesoscale periodicity length, which can be calculated from the average of the reconstructed object autocorrelations (Fourier transform of the measured intensity), which is equivalent to determining the interspacing of the satellite peaks from the diffraction pattern. For the PTO/STO, the 2D calculated periodicity is $(p_x, p_z)$ = (30.3 nm, 25.3 nm). For the $Ca_2RuO_4$ thin film, the autocorrelations show no periodicity in the vertical direction. Therefore, we split the real-space reconstructions into supercells arranged in 1D with a calculated periodicity of $p_x = 18.0$ nm. Then the supercells can be located by finding the local maximum of the cross-correlation between the individual reconstructions and the sinusoidal test function, which is built with the same size as the supercells. Finally, to separate the collected supercells with different patterns, we cluster the supercells using the square-Euclidean-distance k-means++ algorithm[29] with 1000 initializations and a maximum number of iterations of 10000. The clustered supercells are subsequently averaged as the Eigen Pattern of that cluster (see Fig. S2). The code required for generating results shown in Figure 1 is freely available here: https://github.com/ZimingS/Periodic-Textures-Imaging.

**Cryogenic Scanning Transmission Electron Microscopy (cryo-STEM):** Cryo-STEM characterization was performed on a cross-sectional lamella prepared with the standard focused ion beam (FIB) lift-out procedure using a Thermo Fisher Helios G4 UX FIB. Cryogenic HAADF-STEM imaging was performed on a FEI/Thermo Fisher Titan Themis CryoS/TEM with a Gatan 636 double tilt liquid nitrogen cooling holder at 300 kV with a 30 mrad probe convergence semi-angle. For high-precision structural measurements a series of 20 rapid-frame images (~1.6 sec. per frame) were acquired and subsequently realigned and averaged by a method of rigid registration optimized to prevent lattice hops[49] resulting in high signal-to-noise ratio, high fidelity images of the atomic lattice. The c-axis interplanar spacing was extracted from the HAADF-STEM images using the strain mapping technique developed by Smeaton et al.[50] on the 001 peak of the $Ca_2RuO_4$ film. To show a sufficient area of the film two STEM images of overlapping areas of the sample were aligned and montaged together. Unprocessed images and corresponding interplanar spacing maps are shown in Figure S9.

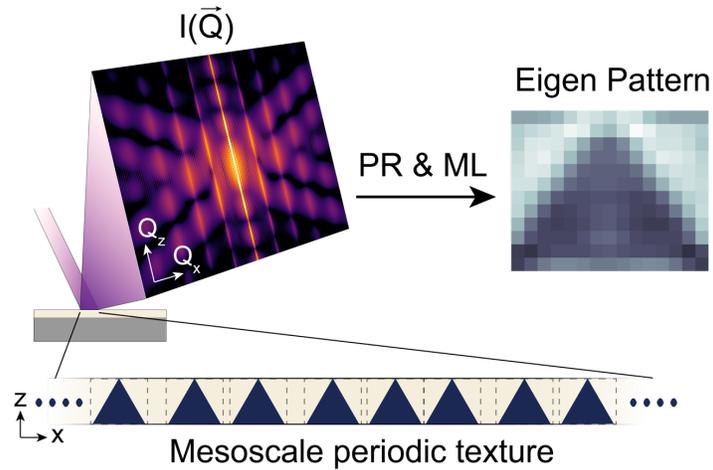

**Figure 1: Real-space imaging of nano-textures in crystalline thin films:** In a Reciprocal Space Map (RSM), the mesoscale periodic ordering is evident from satellite peaks around the Bragg reflection. The diffraction intensity $I(\vec{Q})$ shown here is an average of the squared magnitudes of Fourier transforms of schematic periodic textures with triangular motifs, where the dark area has a constant strain difference compared to the surrounding light area. The motifs are placed with stochastic intervals in the texture exhibiting a translational ordering along the x-axis. Using the $I(\vec{Q})$ simulated as the input, the Eigen Pattern (the triangular structural motif) composing the mesoscale periodic texture can be reconstructed and imaged through a combination of phase retrieval (PR) and unsupervised machine learning (ML).

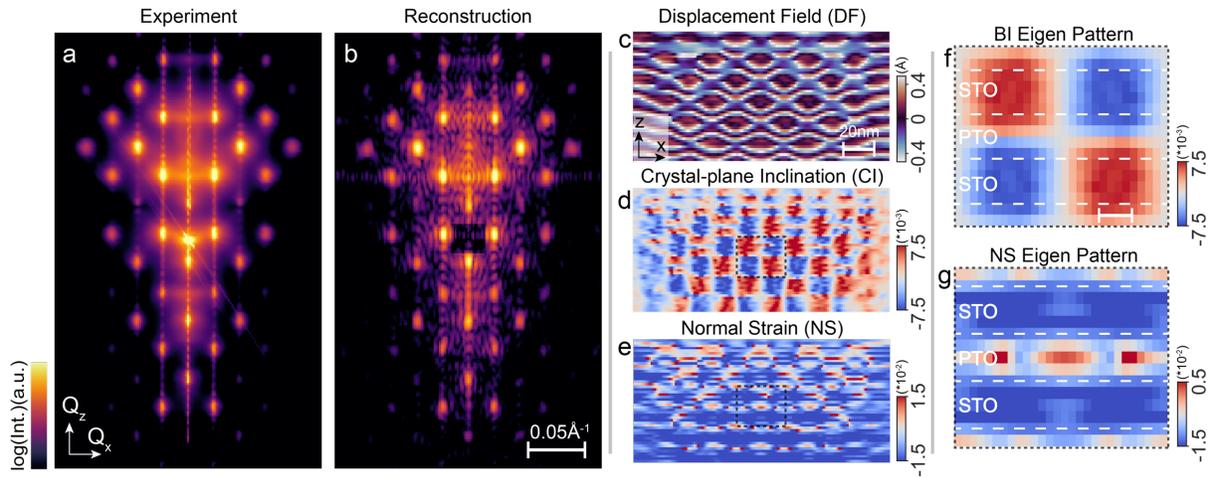

**Figure 2: Imaging of the mesoscale unit cell in the (PbTiO$_3$)$_{16}$/(SrTiO$_3$)$_{16}$ superlattice.** (a) Experimentally measured and (b) reconstructed diffraction pattern near the 004 (pseudo-cubic) peak of (PbTiO$_3$)$_{16}$/(SrTiO$_3$)$_{16}$ superlattice shown on a logarithmic scale. The distinct 2-dimensional periodic features in the measured diffraction pattern are reproduced by the algorithm. The central sharp peak in the measured pattern corresponding to the DyScO$_3$ substrate is removed for the reconstruction. (c-e) Real-space images of the (c) Displacement Field (DF), (d) Crystal-plane Inclination (CI), and (e) Normal Strain (NS) extracted from the data in (a). Alternating layers with a half period lateral translation are observed. (f,g) Real-space images of the (f) CI and (g) NS of the supercell Eigen Pattern composing the entire superlattice. The supercell has a size of 25.3nm and 30.3nm in the vertical and horizontal directions respectively (scale bar shown is 5nm).

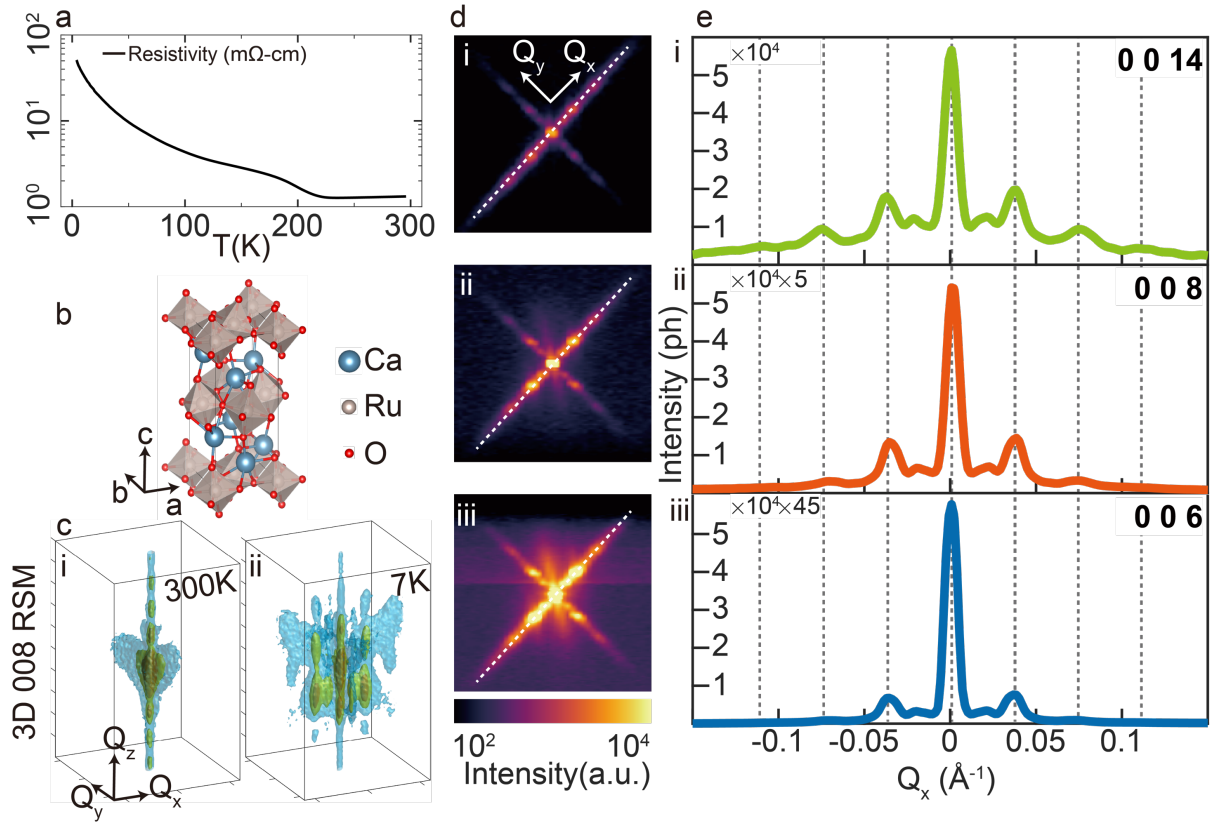

**Figure 3: Mesoscale periodicity in $Ca_2RuO_4$ thin film revealed through X-ray reciprocal space mapping.** (a) Resistivity of the film grown on $LaAlO_3$ (LAO) as a function of temperature. An Insulator-to-Metal Transition (IMT) occurs at around 230K. (b) Structure of the $Ca_2RuO_4$ Pbca lattice and (c) 3D reciprocal space maps (RSM) around the 008 peak of a ~17.5nm $Ca_2RuO_4$ thin film at 300K(i) and 7K(ii) respectively ($Q_z \parallel c \parallel [001]_{LAO}$, $Q_x \parallel a \parallel [110]_{LAO}$, $Q_y \parallel b \parallel [-110]_{LAO}$). The satellite peaks appear at 7K suggesting a temperature induced structural phase transition and the emergence of mesoscale periodicity. (d) Projection of the 7K RSMs around 0014(i), 008(ii), 006(iii) peaks along $Q_z$. (e) Intensity line-profiles of the diffraction pattern along the dashed line in (d). The satellite peaks spacing is constant across multiple Bragg peaks, indicating mesoscale periodicity as the origin of satellite peaks. The broadening of satellite peaks with $Q_x$ suggests distortions in the short-ranged periodicity.

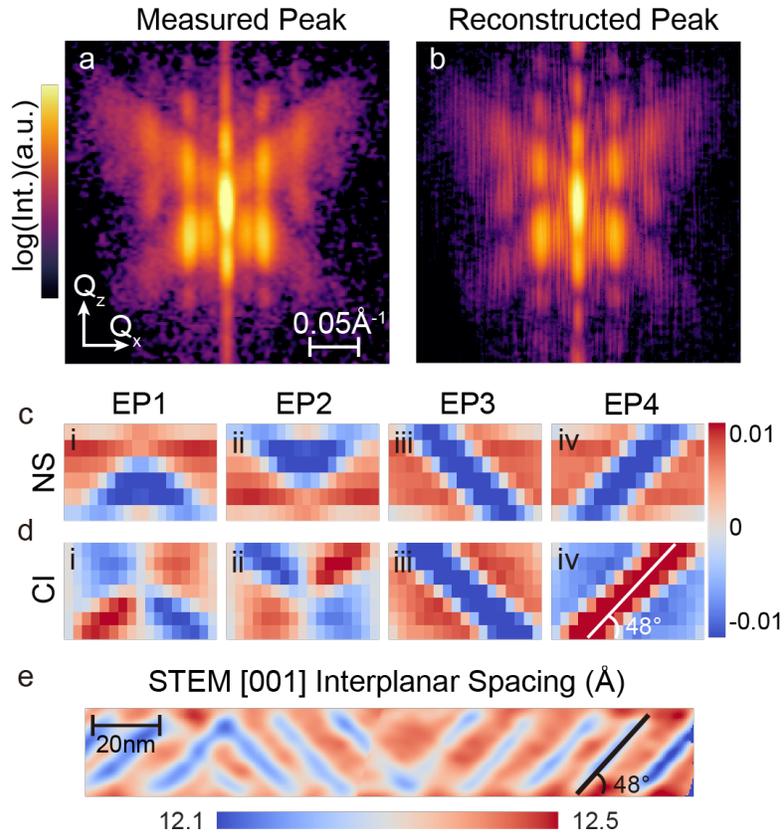

**Figure 4: Imaging of the mesoscale unit cell in the Ca$_2$RuO$_4$/LaAlO$_3$ (CRO/LAO) thin film** (a) Measured and (b) reconstructed diffraction pattern near the 008 peak of CRO/LAO thin film at 7K shown in logarithmic scale. The slice is taken along the dashed white line in Figure 3d. Through reconstruction and clustering, four distinct eigen patterns (EPs) are identified. (c-d) Real-space images of (c) Normal Strain (NS) and (d) Crystal-plane Inclination (CI) in the corresponding EPs. The EPs shown have a size of ~14nm*18nm. (f) Map of the interplanar spacing along [001] extracted from HAADF-STEM images of the cross-section of a ~34nm thick CRO/LAO thin film at ~100K on the [110]$_{LAO}$ zone axis. Both zigzag and diagonal stripe patterns are observed in the image. The average angle of the stripes is close to 48 degrees, which coincides with the angle calculated from the reconstructed eigen pattern.

# Real-space imaging of polar and elastic nano-textures in thin films via inversion of diffraction data: Supplementary Figures

*Ziming Shao[1], Noah Schnitzer[1], Jacob Ruf[2], Oleg Y. Gorobtsov[1], Cheng Dai[3], Berit H. Goodge[4,5], Tiannan Yang[3], Hari Nair[1], Vlad A. Stoica[3,6], John W. Freeland[6], Jacob Ruff[7], Long-Qing Chen[3], Darrell G. Schlom[1,5,8], Kyle M. Shen,[2,5], Lena F. Kourkoutis[4,5], Andrej Singer[1]*

[1]*Department of Materials Science and Engineering, Cornell University, Ithaca, NY 14853, USA*

[2]*Department of Physics, Cornell University, Ithaca, NY 14853, USA*

[3]*Department of Materials Science and Engineering, Pennsylvania State University, University Park, PA 16802, USA*

[4]*School of Applied and Engineering Physics, Cornell University, Ithaca, NY 14853, USA*

[5]*Kavli Institute at Cornell for Nanoscale Science, Cornell University, Ithaca, NY 14853, USA*

[6]*Advanced Photon Source, Argonne National Laboratory, Argonne, IL 60439, USA*

[7]*Cornell High Energy Synchrotron Source, Cornell University, Ithaca, NY 14853, USA*

[8]*Leibniz-Institut für Kristallzüchtung, Max-Born-Straße 2, 12489 Berlin, Germany*

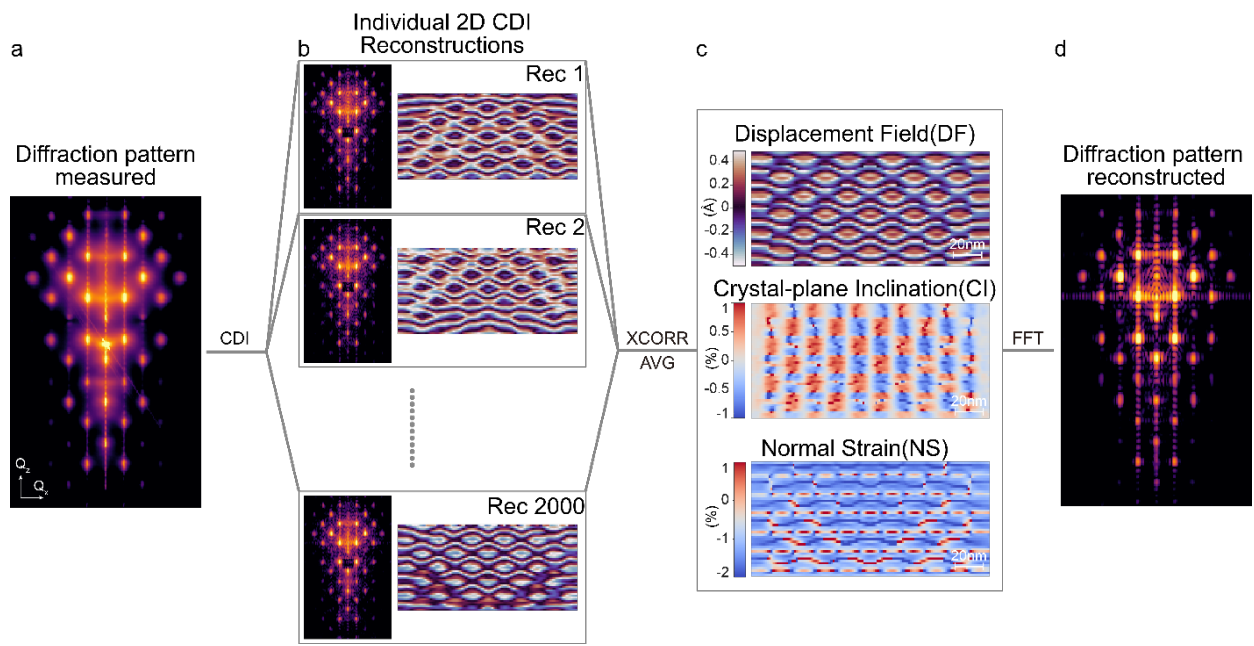

**Figure S1: Phase retrieval and individual reconstructions of (PbTiO$_3$)$_{16}$/(SrTiO$_3$)$_{16}$ (PTO/STO).** (a) Reciprocal Space Mapping around 004$_{pc}$ Bragg peak of (PTO)$_{16}$/(STO)$_{16}$ superlattice upon DSO substrate[1] (b) The calculated diffraction pattern and the corresponding real space phase object of typical independent reconstructions. Most features of the measured pattern are well reconstructed except the sharp DSO substrate peak in the middle, which was intentionally neglected. Periodic-patterned patches with size and location variation are observed within the real-space phase objects. The periodic patches are more likely to be observed in the central region than the edge and corner of the phase objects. The size for the phase object support is 75 nm (vertical) *172 nm (horizontal). (c-d) The result of averaged individual reconstructions. Testing the consistency of the phase retrieval algorithm by averaging over multiple runs of the algorithm initiated by random starting conditions[2–6]: the algorithm is successful when the diffraction pattern of the averaged object agrees with the measured diffraction pattern. Unfortunately, the relative translations in individual reconstructions preclude us from averaging efficiently. In order to find the set of reconstructions with less relative translations, we cross-correlated the individual reconstructions and then average the set of reconstructions with highest mutual correlation. (~100 out of 2000 reconstructions) (c) shows the displacement field (DF), crystal-plane inclination (CI) and normal strain (NS) of the averaged object. (Top to bottom order) All three images show a periodic layered structure that agrees with the PTO/STO superlattice. (d) The corresponding diffraction pattern calculated from the averaged real space phase object. The average intensity successfully reproduces the main features of the measured diffraction data, such as the satellite peaks and the Laue fringes (thickness fringes) in between.

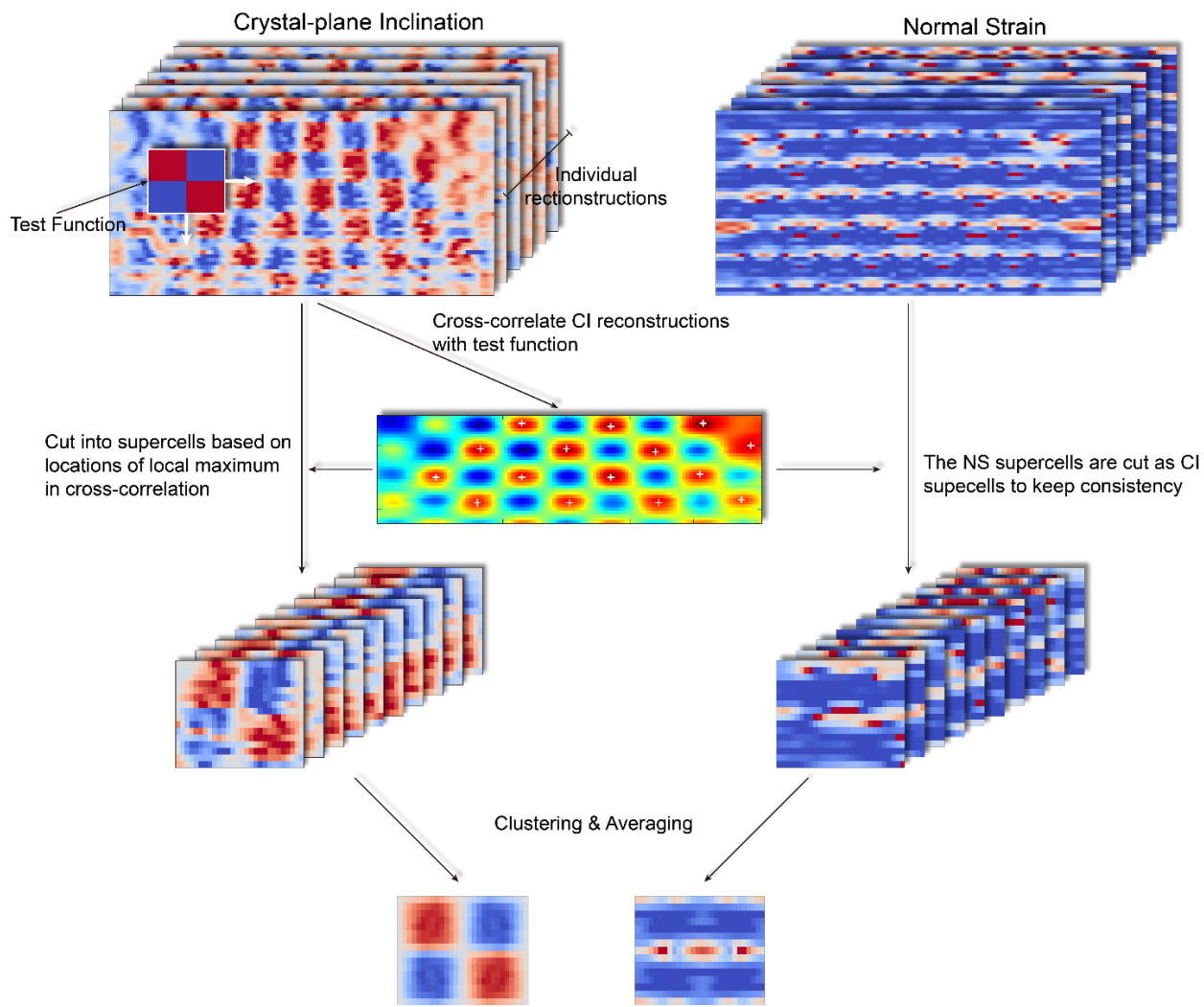

**Figure S2: Flow chart of pattern recognition from a set of individual phase retrieval reconstructions with k-means clustering.**

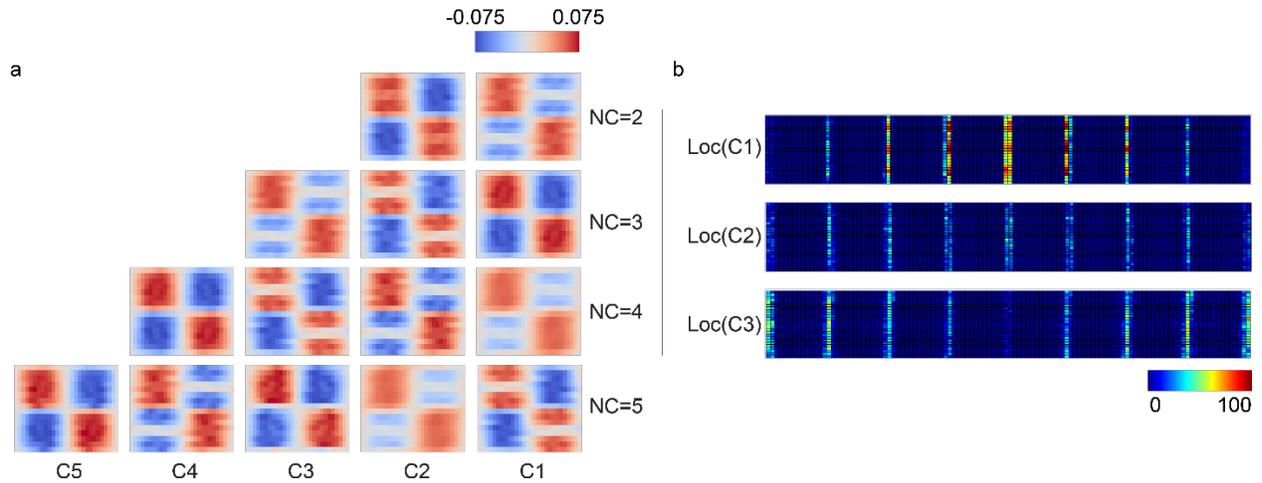

**Figure S3: Clustering results of reconstructions (PbTiO$_3$)$_{16}$/(SrTiO$_3$)$_{16}$ data from reference [1].** Supercells with distinct features are grouped into different clusters by k-means clustering. The characteristic features of one cluster are shown by averaging all the supercells belong to that cluster. Figure (a) shows an average of supercells belonging to every cluster under k-means clustering with different number of clusters (NC). As shown, after NC reaching three, further clustering will not discover additional distinct features other than the three clusters observed in NC=3. For example, when NC=4 both C1 and C2 resemble the C3 of NC=3, and the same situation can also be observed in NC=5. (b) 2D histograms of the supercells positions of each cluster when NC=3. Cluster C1 has the largest population and the distribution of C1 supercells concentrates in the center of the reconstructed object where the periodic patches mostly observed. Therefore, C1 is considered as the Eigen pattern that composing the mesoscale periodicity. Meanwhile, the averaged patterns of C2 and C3 resemble C1 except the horizontal bisection of the diagonal blocks. Because C2 and C3 have smaller population compared to C1 and prone to appear at the reconstruction edges, the reliability of the bisection feature is skeptical. Moreover, in all the three histograms, the lateral positions of the supercells are well defined and evenly spaced, while a position uncertainty is noticed in the vertical direction, which reemphasizes the necessity to perform averaging over supercells instead of the entire reconstructed object.

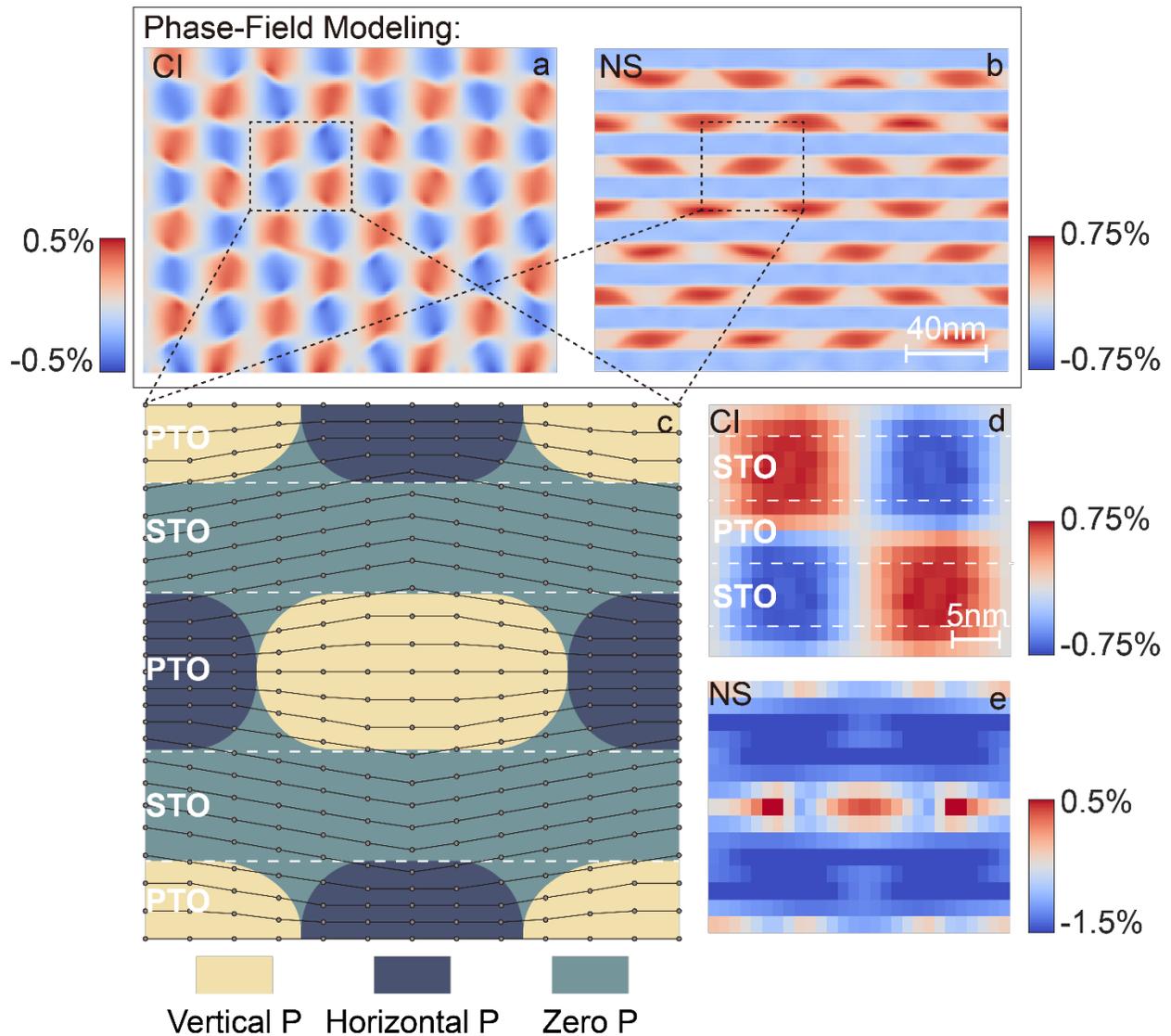

**Figure S4: Comparison reconstruction results with Phase-field modeling** Crystal-plane inclination map (a) and normal strain map (b) calculated from phase-field modeling result. Both maps show 2D periodicity. The corresponding supercells are indicated by dashed boxes. (c) Schematic of the polarization and lattice in the supercell. The three colors represent three different local polarization statuses: polarized vertically, polarized horizontally, and zero polarization. As the polarization is correlated with the distortion of the PTO unit cells, the local vertical lattice spacing can be inferred as the black dots and lines shown in the schematic. The region with horizontal polarization has smaller vertical lattice spacing than the region with vertical polarization. And because of the local 'shrink' and 'expand' of the PTO layer, inclination will occur in the adjacent STO layer lattice accordingly. The inclination (d) and normal strain supercells (e) reconstructed by our method agree well with the results from the phase field modeling. Both show a 2d checkerboard pattern in the CI and a layered pattern in NS. Moreover, the CI supercells from the phase field modeling also show a slightly stronger inclination around the horizontal interface of the opposite blocks, appearing a bisectional feature similar to what we observed in one of the clusters of the CI supercells (Figure S3a).

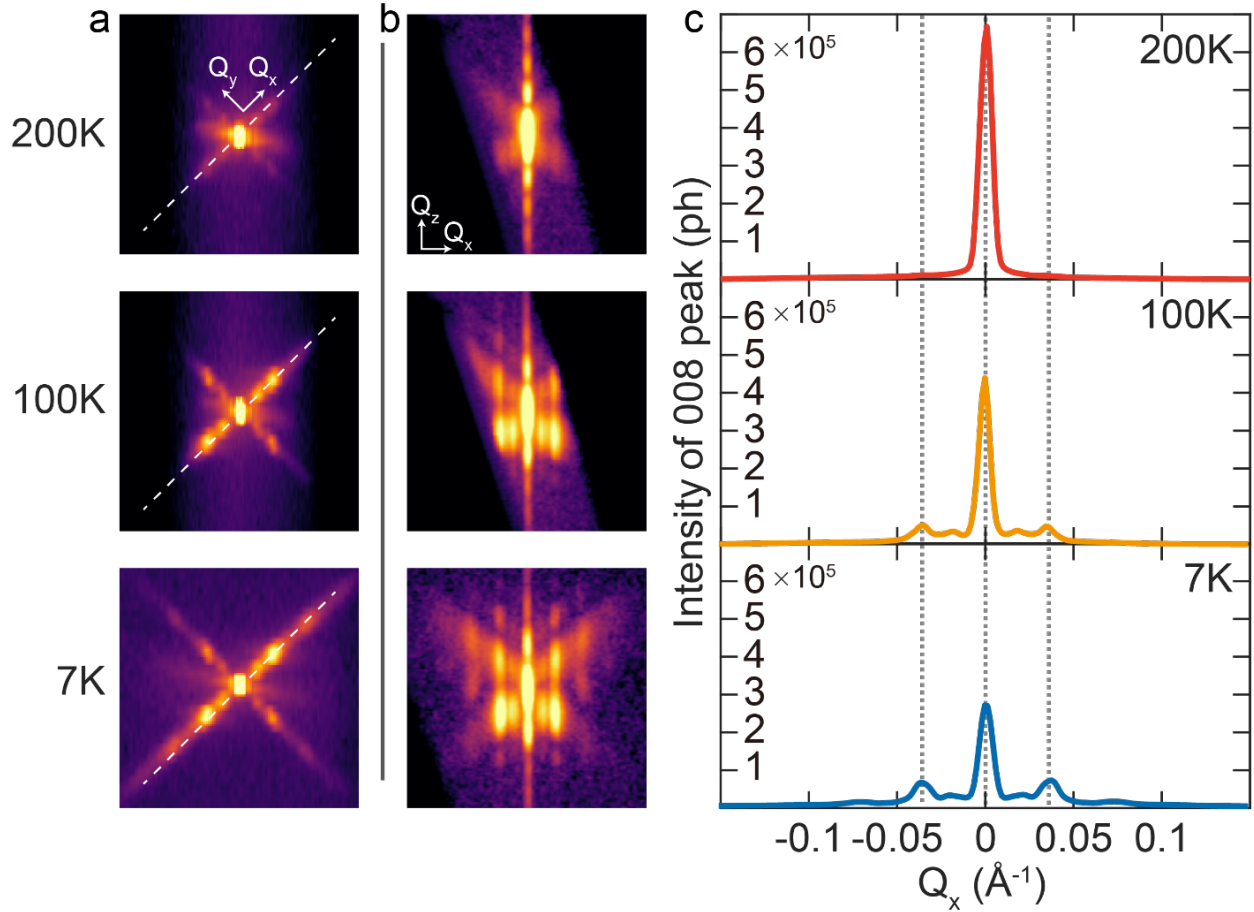

**Figure S5: Diffraction patterns under various temperatures** The RSMs around $Ca_2RuO_4$ (CRO) 008 Bragg peak measured with area detector using θ-scan, under 200K, 100K and 7K. (a) Projection of the diffraction pattern along $Q_z$. (b) The corresponding 2D slices and (c) line profiles along the diagonal white dashed lines in (a). With the decrease of the temperature, the satellite peaks emerge gradually. And from the line profiles of the dashed lines as well as the projections, the spacing between the satellite peaks appear to be constant indicating the period of the short-range order doesn't seem to change as a function of temperature. The spacing between the satellite peaks is 0.36 $nm^{-1}$, corresponding to a period length of ~17.5 nm. (The θ-scan under 200K and 100K have a smaller θ range than the 7K and only the central part of the entire peak is recorded).

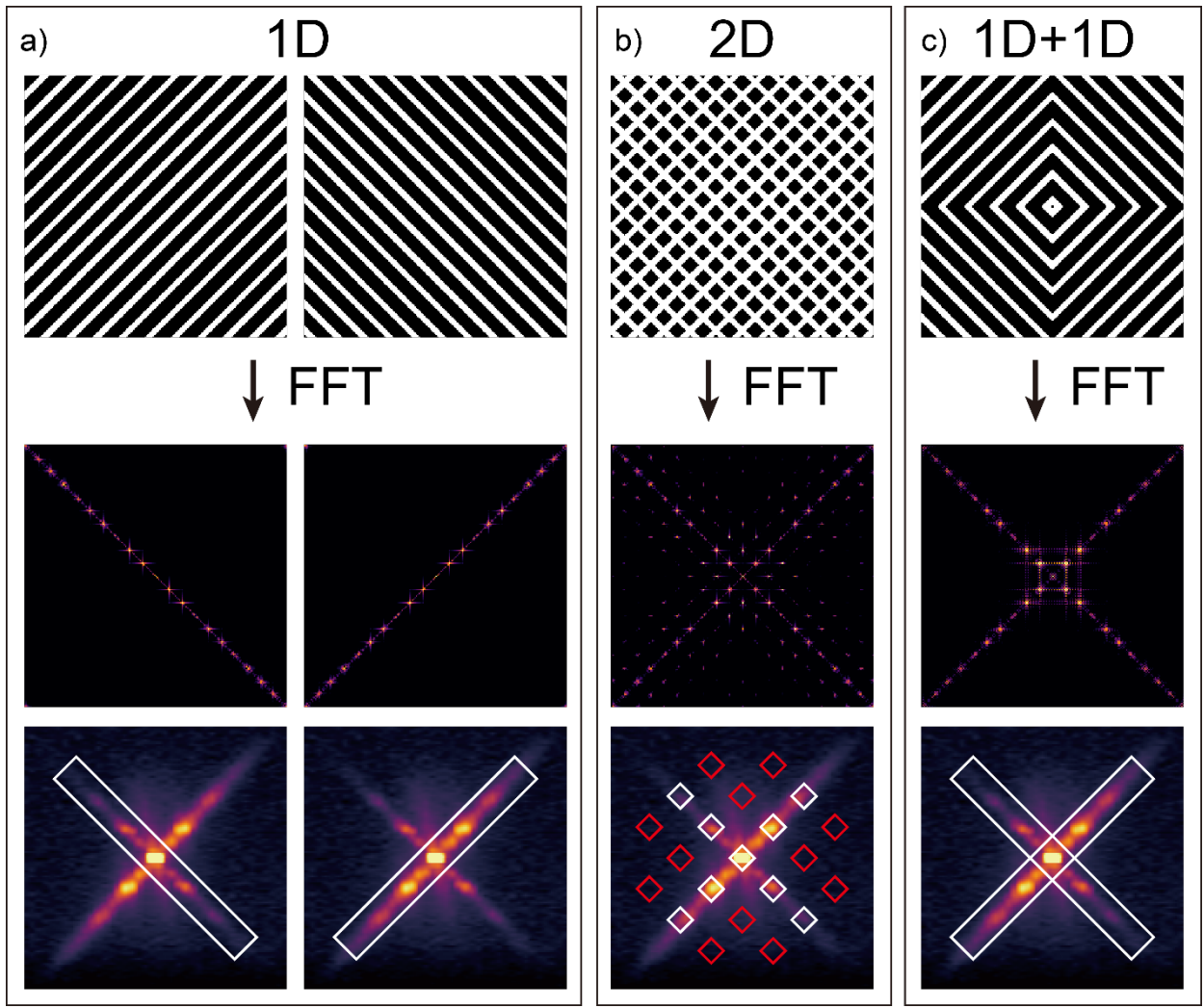

**Figure S6: Indication of spatially separated 1D periodic domains in Ca$_2$RuO$_4$ thin film** (a) A structure with 1D periodicity will lead to a 1D periodic pattern with the same periodicity orientation after Fourier Transform (FT). (b) A tweed structure will lead to a 2D periodic pattern after FT. In the measured diffraction pattern of CRO, satellite peaks only emerge on the diagonals (white boxes). And no peaks were observed at the off-diagonal positions (red boxes), which are expected from a 2D periodic tweed structure. (c) A spatially separated 1D periodic structure with orthogonal orientation will lead to a cross shaped diffraction pattern as observed in the diffraction pattern of CRO. Moreover, an evident intensity difference can be observed between the two diagonals: The integral intensity of the diagonal along top-right to bot-left approximately three times than the other diagonal. Therefore, it seems that each diagonal streak represents spatially separated domains with 1D periodicity oriented along one of the orthogonal directions. And the relative intensity difference of the two streaks comes from the volume disparity of the two types of domains.

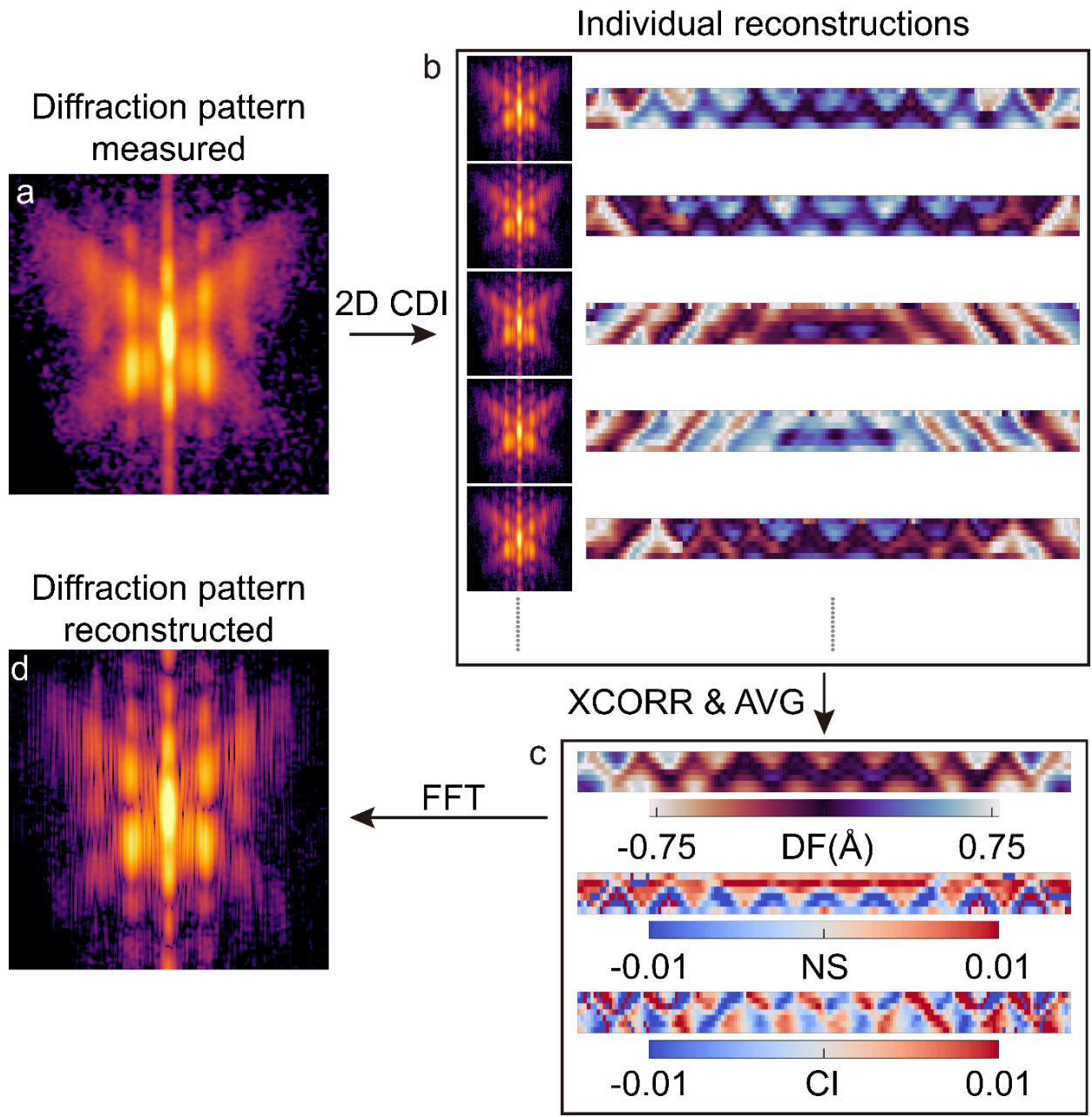

**Figure S7: Phase retrieval and individual reconstructions of CRO.** (a) RSM measured around CRO 008 Bragg peak (b) The calculated diffraction pattern and real space phase object of individual reconstructions calculated using phase retravel algorithms. As shown, the reconstructed patterns share identical features as the measured peak and all the corresponding real space phase objects show some level of short-range order. The corresponding real space phase objects may vary significantly. Some show zigzag patterns, and some appear to contain diagonal stripes. The support for the real space phase object used here is 14 nm (vertical) *170 nm (horizontal). (c) The result of averaged individual reconstructions. (d) The diffraction pattern of the averaged real space phase object.

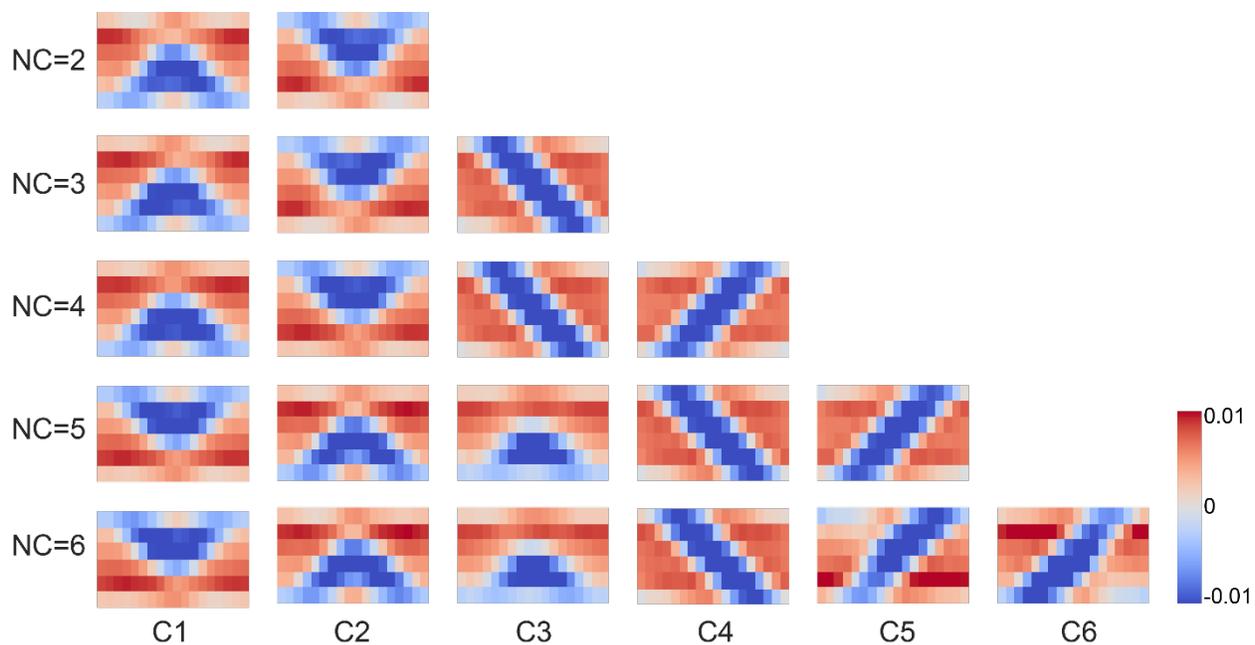

**Figure S8: Clustering results of reconstructions of $Ca_2RuO_4$ data.** Average of normal strain (NS) supercells after k-means clustering with different number of clusters (NC). Compared to the method shown in S2, 1D instead of 2D cross-correlation is used for this reconstruction since the periodicity only exist in one direction. As shown, k-means clustering with four clusters result in four distinct eigen patterns, while clustering with more clusters will result in redundant eigen patterns. For example, when NC=5, C2 and C3 share a similar triangular feature; when NC=5, two pairs C2/C3, C5/C6 both show a similar feature. NC=4 is selected in the k-means clustering to avoid unnecessary excessive clustering.

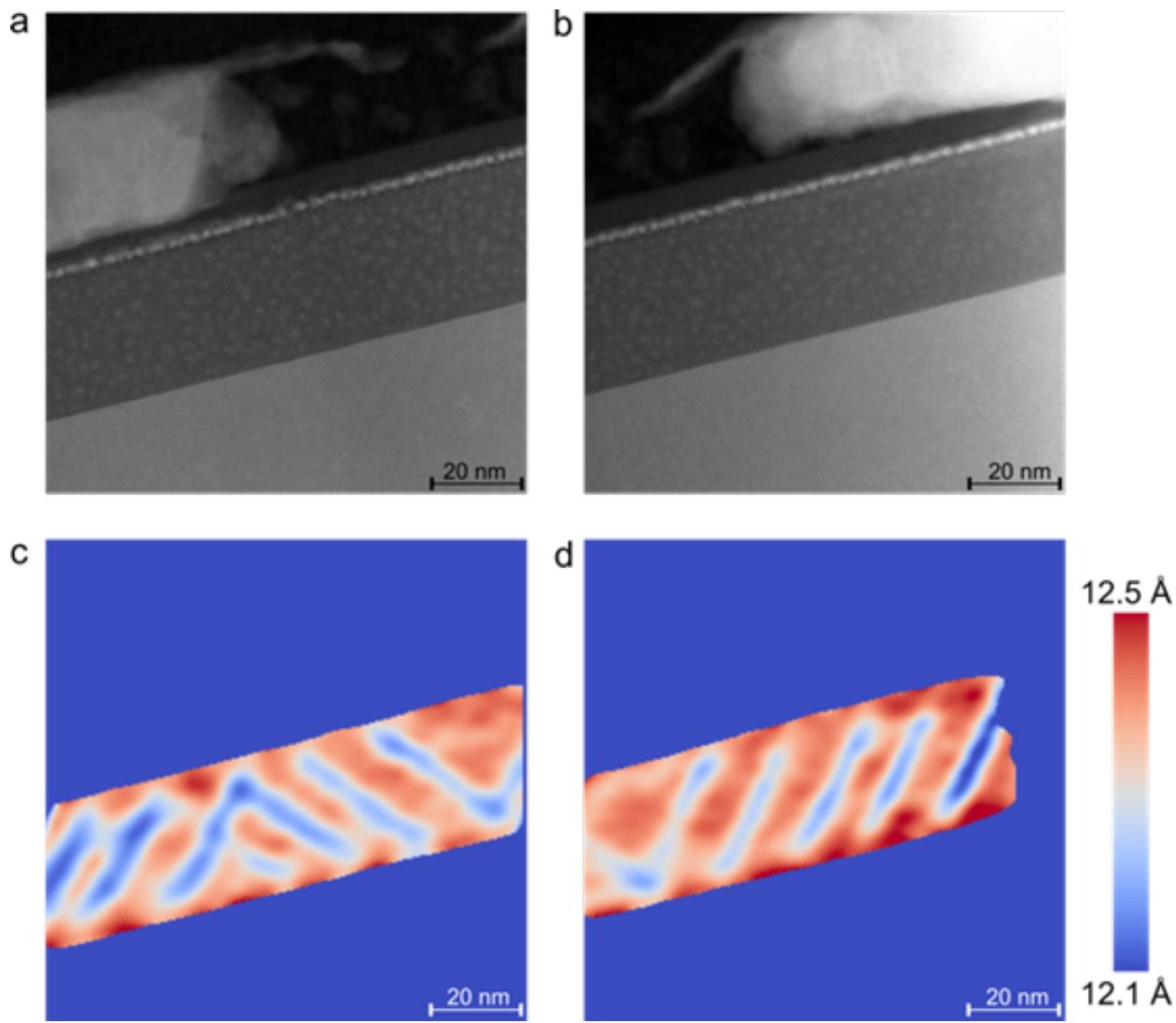

**Figure S9: Cryogenic Scanning Transmission Electron Microscopy (cryo-STEM) results.** (a,b) Unprocessed HAADF-STEM images of the CRO film cross-section (c,d) Corresponding [001] interplanar spacing maps (see methods in the main text).

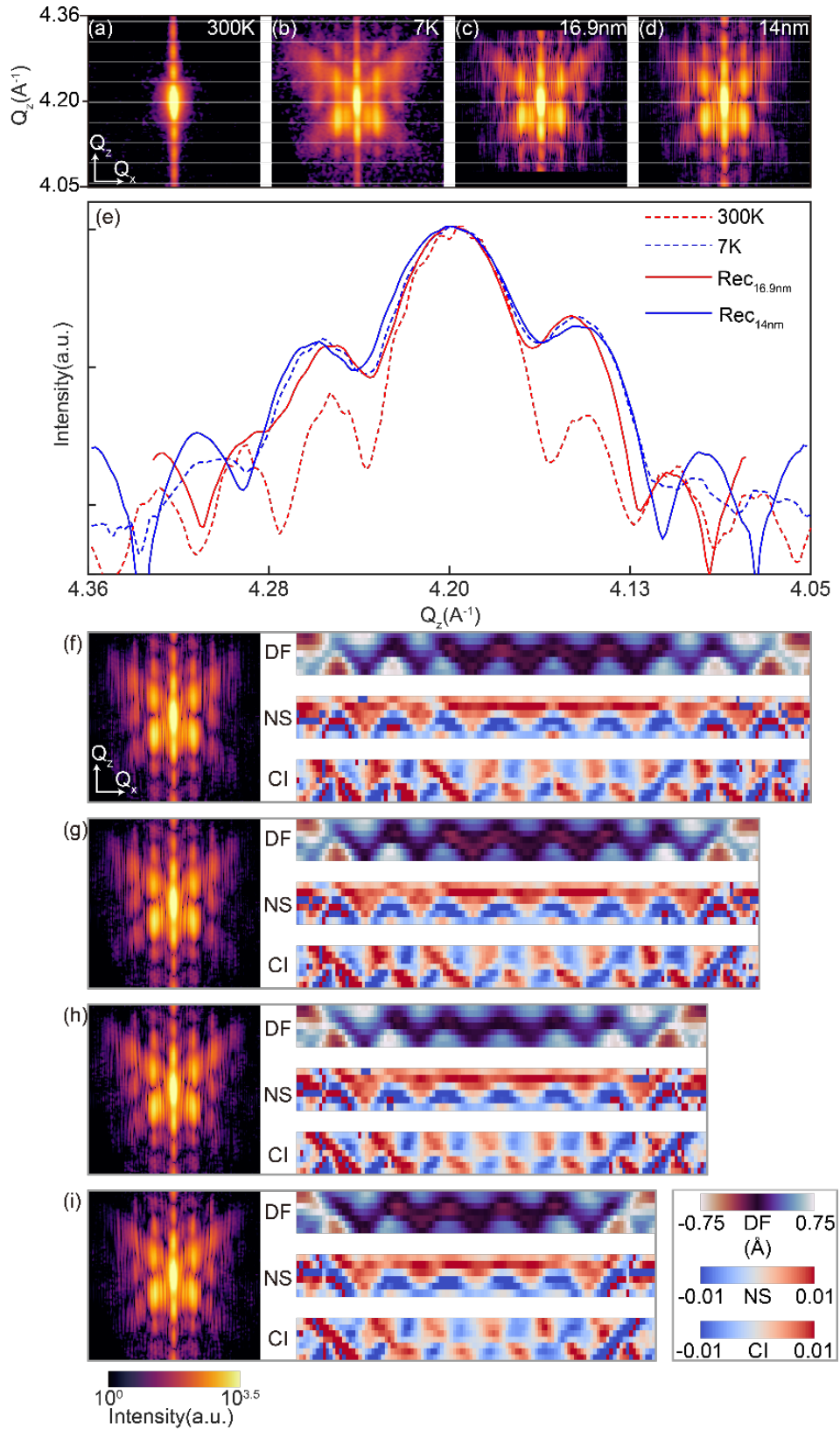

**Figure S10: Reconstruction with various support shapes** (a, b) Measured X-ray diffraction pattern in the plane of $Q_z[001]_{LAO}$-$Q_x [110]_{LAO}$ under (a) 300K and (b) 7K, respectively. A difference of the thickness fringes can be noticed by comparing the two

patterns (white grid is set according to the thickness fringes minimum positions at 300K). The 300K fringes indicate a film thickness of 16.9 nm, while the 7K fringes indicate a thickness of 14 nm. The cause of the disparity of the thickness fringes under different temperatures is unclear. One plausible explanation is the illuminated region upon the thin film drifted during the dramatic change of temperatures because of the thermal expansion of the sample holder. And a thickness change of ~3nm is possible on CRO/LAO film as observed in the STEM. (c, d) The real space pixel size in the phase retrieval algorithm is directly related to the reciprocal space range. By tuning the input reciprocal space image size, we perform reconstructions with support thickness of 16.9 nm and 14nm. The reconstructed diffraction patterns with support thickness of (c) 16.9nm and (d) 14 nm. (e) The comparison of the thickness fringes of measured and reconstructed patterns. An obvious difference exists between the two measured fringes especially at the off-center regions. The fringe of the 16.9 nm reconstruction agrees with the measured 300K fringe, and the 14 nm reconstruction fringe agrees with the measured 7K fringe. Therefore, the support thickness is directly related to the thickness fringes of the reconstructed pattern and a support thickness of 14 nm is selected to reconstruct the data measured at 7K. Compared to the support thickness, the horizontal size of the support has less effect on the reconstruction. (f)-(i) Reconstructions with thickness of 14nm and various horizontal support size of 169.5nm, 150.5nm, 133.7nm, 116.8nm. All the reconstructed diffraction patterns well retained the features of the measured data and resemble each other. And the width of the fine vertical lines within the reconstructed pattern is inversely related to the support horizontal size, as the a 'thickness fringes' analogue in the horizontal direction. Although the reconstructions are with different support size, similar repeating patterns are observed in all the reconstructed real space object with different number of periods.

# Sample 2D Phase Retrieval Code
## Define support and initialize input

```
%data is a 2D matrix of diffraction intensity, with the center of the
%bragg peak aligned to the center of the matrix
srhor=80; %Support horizontal size
srver=4;  % Support vertical size
Simnum=1;%number of random starts/simulations;
Collector=zeros([size(data),Simnum]);%initialize collector for individual reconstruction results

support=zeros(size(data));
[sx,sy]=meshgrid(-size(data,2)/2:size(data,2)/2-1,-size(data,1)/2:size(data,1)/2-1);
support((abs(sx)<srhor)&(abs(sy)<srver))=1;%real-space constraint
G0=sqrt(data);%Calculate Amplitude from Diffraction Intensity(data)
G0=ifftshift(G0);
support=ifftshift(support);
```

## Iterative Phase Retrieval

```
for nn=1:Simnum
    rng shuffle
    gk=exp(1j*2*pi*rand(size(data)));% random start
    for n=1:1999 %number of iteration in one reconstruction with random start
        G0c=G0;
        if mod(n,100)<60
            gk=OOPCSYM(gk,G0c,support,0.6);
        elseif mod(n,100)<80
            gk=OOPC(gk,G0c,support,0.8);
        else
            gk=OOPC(gk,G0c,support,0.98);
        end
    end
    clear Gk Gkp
    Collector(:,:,nn)=gk;
end
support=fftshift(support);
G0=fftshift(G0);
```

## Custom Output-Output algorithm

```
function gkk=OOPC(gk,G0,support,beta)
Gk=ifft2(gk);
Gkp=G0.*exp(1j*angle(Gk));%Fourier space constrain
gkk=fft2(Gkp);
gkk=exp(1j*angle(gkk));
gkk=support.*gkk+(1-support).*(gkk-beta*gkk);%output-output
end
```

## Custom Output-Output algorithm with real-space SYM constraint

```
function gkk=OOPCSYM(gk,G0,support,beta)
Gk=ifft2(gk);
Gkp=G0.*exp(1j*angle(Gk));%fourier space constrain
gkk=fft2(Gkp);
gkk=exp(1j*angle(gkk));
gkk=SYM(gkk);%real space symmetry constraint
gkk=support.*gkk+(1-support).*(gkk-beta*gkk);%output-output
end
```

## Real Space Symmetry Constraint

```
function gkk=SYM(gkp)
gkk=exp(1j*(angle(gkp)+fliplr(angle(gkp)))/2);
end
```